\documentclass{tPHM2e}
\usepackage{graphicx}
\usepackage[]{epsfig}

\newcommand{\beq}{\begin{equation}}
\newcommand{\eeq}{\end{equation}}
\newcommand{\beqd}{\begin{displaymath}}
\newcommand{\eeqd}{\end{displaymath}}
\newcommand{\beqa}{\begin{eqnarray}}
\newcommand{\eeqa}{\end{eqnarray}}

\newcommand{\sign}{{\rm sign}}

\newcommand{\comment}[1]{}

\newcommand{\ttau}{\tilde{\tau}}
\newcommand{\tlet}{\tilde{t}}

\begin{document}

\title{The Glass Crossover from Mean-Field Spin-Glasses to Supercooled Liquids}

\comment{
\author{Tommaso Rizzo$^{1,2}$}
\affiliation{
$^1$ ISC-CNR, UOS Rome, Universit\`a "Sapienza", PIazzale A. Moro 2, \\
$^2$ Dip. Fisica, Universit\`a "Sapienza", Piazzale A. Moro 2, I-00185, Rome, Italy \\
I-00185, Rome, Italy}
}

\author{
\name{Tommaso Rizzo\textsuperscript{a,b}$^{\ast}$\thanks{$^\ast$Corresponding
author. Email: tommaso.rizzo@cnr.it}}
\affil{\textsuperscript{a}ISC-CNR, UOS Rome, Universit\`a "Sapienza", Piazzale A. Moro 2, I-00185, Rome, Italy; \textsuperscript{b}Dip. Fisica, Universit\`a "Sapienza", Piazzale A. Moro 2, I-00185, Rome, Italy}
 }

\maketitle

\begin{abstract}
Stochastic-Beta-Relaxation (SBR) provides a characterization of the glass crossover in discontinuous Spin-Glasses and Supercoooled liquid. Notably it can be derived through a rigorous computation from  a dynamical Landau theory. In this paper I will discuss the precise meaning of this connection in a language that does not require familiarity with statistical field theory. I will discuss finite-size corrections in mean-field Spin-Glass models and loop corrections in finite-dimensional models that are both described by the dynamical Landau theory considered. Then I will argue that the same Landau theory can be associated to supercooled liquid described by Mode-Coupling-Theory invoking a physical principle of time-scale invariance. 
\end{abstract}

\section{Introduction}

The dramatic slowing down of dynamical relaxation of many super-cooled liquids is characterized by the existence of a crossover temperature where there is a change in the growth rate of the relaxation time from power-law like to a more pronounced exponential growth. The nature and origin of this crossover is a key issue in glass theory. Some authors believe that all the physics above $T_g$ can be understood solely in terms of this crossover without advocating the presence of true dynamical singularity at a temperature $T_K<T_g$.

Understanding the crossover is also an important step in order to make theoretical progress.
From the theoretical point of view the most successful first-principles and quantitative theory of the supercooled state is Mode-Coupling-Theory (MCT) \cite{Gotze09}. However its validity is limited to a range of temperature considerably higher than $T_g$ and the reason is that MCT mistakenly predicts that in place of the crossover there is a true dynamical singularity characterized by a diverging relaxation time which is not at all observed in real systems.

Stochastic-Beta-Relaxation (SBR) is model of the glass crossover that I have introduced in a recent
 publication \cite{Rizzo14}.  The model describes the time evolution of the density-density correlator of the liquid in the $\beta$-regime, {\it i.e.} the time window where it stays near a plateau.
Within MCT \cite{Gotze09} the density-density correlator has the following behavior in the $\beta$ regime:
\beq
{\bf \Phi}(k,t)={\bf F}(k)+G(t) \, {\bf H}(k)\, 
\label{scalfor}
\eeq
where the bold character accounts for the case of mixtures of particles in which the correlator is a matrix.
The function $G(t)$ obeys the well-known MCT equation for the critical correlator:
\begin{displaymath}
\sigma=-\lambda \, G^2(t) +{d \over dt}\int_0^t G(t-s)G(s)ds
\end{displaymath}
where the separation parameter $\sigma$ is negative at high temperatures (low pressures) and vanishes at the MCT singularity. The above equation predicts that the correlator will eventually leave the plateau  above the critical temperature ($\sigma<0$) while it will remain a in a glassy state below the critical temperature ($\sigma>0$). 

SBR can be viewed as an extension of the MCT equation for
the critical correlator with random fluctuations of the separation parameter.
According to it in the $\beta$ regime, equation (\ref{scalfor}) continues to hold and only the epxression of the critical correlator $G(t)$ is different.
One must consider a field $g(x,t)$ that is a local version of the correlator and that obeys the following equation:
\beq
\sigma  + s(x) =- \alpha \nabla^2 \, g(x,t)-\lambda \, g^2(x,t)+{d \over dt}\int_0^t g(x,t-s)g(x,s)ds
\eeq
where the field $s(x)$ is a {\it time-independent} random fluctuation of the separation parameter, Gaussian and delta-correlated in space:  
\beq
[s(x)]=0\, ,\ [s(x)s(y)]=\Delta \sigma^2  \, \delta (x-y) 
\eeq
The above equation must be solved with a (conventional) uniform initial condition $g(x,t) \propto 1/t^a$ at small times. 
The  total correlator is obtained as the integral over space averaged over the random fluctuations:
\vspace{10px}
\beq
G(t) = {1 \over V}\int[ g(x,t)] dx
\eeq

The model is attracting considerable interest as it offers a consistent solution to the problem of the  MCT singularity while lacking the drawbacks of the many proposals appeared earlier in the literature. Indeed as discussed in \cite{Rizzo14} within SBR the unphysical singularity predicted by MCT is avoided and replaced by a dynamical crossover from relaxational to activated-like dynamics.
Further study of the SBR equations is unvealing a rich phenomenology and a rather non-trivial characterization of the qualitative and quantitative changes occurring at the glass crossover \cite{Rizzo14b,Rizzo15}.
These include notably a change in the spatial structure of dynamical fluctuations characterized by the appearance of strong Dynamical Heterogeneities and violations of the Stokes-Einstein relationship \cite{Rizzo14b}.
Another non-trivial feature is that the increase of the relaxation time and dynamic susceptibility is accompanied by a {\it decrease} of the dynamical correlation length below the crossover temperature \cite{Rizzo15}, in contrast to the classic Adams and Gibbs \cite{Adams65} view that dynamic slowing down is intrinsically associated to an ever increasing correlation length which is the size of the Cooperatively Rearranging Regions.

In this paper I want to discuss another important aspect of SBR, {\it i.e.} the fact that it is not a phenomenological theory: in particular the random fluctuations of the temperature that are its main features are not just a nice idea that comes out of nowhere but they are the result of a rigorous and non-trivial calculation.
It is important to appreciate this point that put SBR on a different ground in the field of glass theory which is full of theories that are either admittedly phenomenological or otherwise very speculative.
In a few words the calculation shows that there is an equivalence between a dynamical effective theory {\it a la} Landau (called glassy critical theory (GCT) below) on one side and SBR on the other side. The explanation of this connection was sketched in \cite{Rizzo14} in a way accessible for people familiar with statistical field theory and the field theoretical approach to critical phenomena.
While the actual derivation is rather technical and still unpublished, one does not need to master all the technical details in order to understand it physically.
In the following I will discuss this connection more clearly and in physical terms in a way hopefully accessible also to the general reader in the field who is not necessarily familiar with statistical field theory.
I will start with the discussion of finite-size corrections in mean-field Spin-Glass (SG) models, showing that they are indeed given by SBR, later I will discuss the relevance of the equivalence for finite-dimensional SG, and finally I will discuss the relevance of SBR for supercooled liquids through MCT by invoking the physical principle of time-scale invariance.

\section{Effective Theories and Stochastic-Beta-Relaxation}

\subsection{Finite-Size Corrections}

In order to be definite I will consider a SG models of $N$ spins $s_i$ that interact through a quenched random Hamiltonian. The analog of the dynamic correlator in supercooled liquids is 
the quantity
\beq
C(t)=\overline{\langle s_i(0)s_i(t)\rangle}
\eeq  
where the angle bracket means average over the dynamics and the overline means average with respect to the quenched disorder.
For large values of $N$ this quantity can be written as an expansion in powers $1/N$ (The size of the system) that give the finite-size corrections to the thermodynamic limit result:
\beq
C(t)=C_0(t)+{1\over N}C_1(t)+{1\over N^2}C_2(t)+\dots
\label{Cexp}
\eeq
Now it is well known that there is class of SG models called one-step Replica-Symmetry-Breaking (1RSB) where the correlator in the thermodynamic limit exhibits a MCT-like singularity \cite{Kirkpatrick87c,Crisanti93}, in other words it exists a critical temperature (called $T_d$ in SG literature and $T_c$ in MCT literature) close to which the correlator develops a plateau at some value $C_p$:
\beq
C_0(t) \approx C_p+ \delta C_0(t)
\eeq 
with $\delta C_0(t)\ll 1$ in a certain time window called the $\beta$ regime.
In particular {\it below} the critical temperature the function $C_0(t)$ in the infinity time limit does not decays to the paramagnetic/liquid value but remain blocked at the  therefore if we consider the quantity
\beq
q_i=\lim_{t\rightarrow \infty} C_i(t)
\label{qi}
\eeq 
we can study the finite size corrections to the plateau value in the glassy phase at different temperatures:
\beq
\lim_{t \rightarrow \infty} C(t) \equiv q = q_0+{1\over N}q_1+{1\over N^2}q_2+\dots
\eeq
Precisely at the critical temperature we have by definition $q_0=C_p$ while $q_0 > C_p$ at lower temperatures. Note that the $q_i$'s are only defined below the critical temperature but we have considerably simplified the problem by moving from dynamics to statics, as we will see however in order to 
completely solve the problem a purely static is not sufficient.
In the context of fully connected  mean-field Spin-Glass models one can study the various $q_i$'s explicitly and it turns out that they are singular as a function of $\tau\equiv T-T_c$.  The leading order $q_0$ has itself a square root singularity while each finite size correction is divergent.
The nature of these divergences is of the form:
\beq
q_i=\tau^{-2 i+1/2} a_i+o(\tau^{-2 i+1/2})
\label{qlo}
\eeq
One can show on general grounds  that the above form is correct for all values of $i$ ({\it i.e.} at all orders) but the {\it actual} values of the coefficients $a_i$ requires an explicit computation and typically is limited to the first few orders.

Therefore we see that each correction is divergent and furthermore higher order corrections that in principle are smaller in the thermodynamic limit because they are associated to a larger values power of $1/N$ have instead a higher degree of divergence.
On the other hand we know that there is no critical temperature in a finite size system and therefore we expect that somehow the perturbative expansion (that corresponds to first taking the limit $N \rightarrow \infty$ and afterwards the limit $\tau \rightarrow 0$) must break down at fixed $N$ if we are close enough to $\tau=0$.
One can easily see that the above series can be rearranged in such a way  that at {\it fixed} value of $N$ there is a range of values of $\tau$ where all corrections become of the same order, indeed if we write
\beq
\tau=N^{-1/2} \tilde{\tau}
\eeq
we have 
\beq
q=C_p+N^{-1/4}\sqrt{\tilde{\tau}}\left(a_0+{a_1 \over \tilde{\tau}^2}+{a_2 \over \tilde{\tau}^4}+\dots\right)+o(N^{-1/4})
\eeq
where the $o(N^{-1/4})$ corrections to the above expression come from the corrections term to the leading-order divergences of the $q_i$ {\it i.e.} the $o(\tau^{-2 i+1/2})$ terms in (\ref{qlo}).
Therefore we would guess that at the critical temperature $\tau=0$ we have finite-size corrections to the plateau value that are $O(N^{-1/4})$ {\it i.e.} larger than below the critical temperature where they are $O(1/N)$. In particular we could write
\beq
q=C_p+{a_{scal}(0) \over N^{1/4}}+o(N^{-1/4}) \ , \  (\tau=0)
\label{fiq}
\eeq
where the scaling function $a_{scal}(\tilde{\tau})$ is defined as
\beq
a_{scal}(\tilde{\tau})\equiv \sqrt{\tilde{\tau}}\left(a_0+{a_1 \over \tilde{\tau}^2}+{a_2 \over \tilde{\tau}^4}+\dots\right)
\eeq
and one should resum the all series in order to get a finite value at $\tilde{\tau}=0$.
In similar situations in mean-field systems  one can explicitly evaluate only a few number of the first coefficients $a_i$ and then face the problem of using them in order to guess somehow the value of $a_{scal}(0)$. However before proceeding to the discussion of the coefficient $a_{scal}(0)$ let us notice that expression (\ref{fiq}) is troublesome.
Indeed we expect that a system of finite size is ergodic and therefore we would expect finite size effects to destroy the transition in the sense that the correlator should always decay to zero (or the value corresponding to paramagnetic/liquid phase). Correspondingly we would have $C(\infty) \neq C_p$ and the all expansion should break down. 
Note however that the perturbative expansion of $a(\tilde{\tau})$ is perfectly defined, we can compute its coefficients $a_i$ to any desidered order (with increasing computational effort) and we would never see any problem that could only manifest itself as a non-perturbative effect.

We are now in position to discuss the first important breakthrough obtained in recent years in \cite{Franz11b}.  Although the original paper addresses the problem from a somewhat different perspective and language one of its main points is the statement and proof of following theorem:
\begin{itemize}
\item {\it The coefficients $a_i$ are the same that are generated by the expansion of the following expression}
\beq
a_{scal}(\ttau) \sim [\sqrt{\ttau+h}]
\eeq 
{\it where the square brackets mean average over $h$ that is random Gaussian variable with zero mean and non zero variance $\Delta \sigma^2$.}
\end{itemize}
In order to determine the $a_i$'s one can factorize the $\sqrt{\ttau}$, expand the factor $[\sqrt{1+h/\ttau}]$ in powers of $1/\ttau$ and then average over the $h$'s.
Note that subtlety of the above result: it does not give us an explicit expression for the coefficients $a_i$, (we still have to compute them, although the computation is considerably simpler) but it tells us that they are the {\it same} that would come out expanding a different expression.   
On the other hand we are  not really interested in the $a_i$'s {\it per se} but rather in their resummed expression that should give us the function $a_{scal}(\ttau)$. At this stage it seems natural to identify 
$a_{scal}(\ttau)$ with $[\sqrt{\ttau+h}]$ {\it i.e.} to say that they are truly equal, not only order by order in perturbation theory; This identification however is troublesome because we see that the quantity $[\sqrt{\ttau+h}]$ is only defined as an asymptotic expansion in powers of $1/\ttau$, if we try to compute it for a given value of $\ttau$ we run into the problem that $h$ is a Gaussian variable and therefore the integral
\beq
\int_\infty^\infty{dh \over \sqrt{2 \pi \Delta \sigma}}\exp[-h^2/2\Delta\sigma]\sqrt{\ttau+h}
\eeq 
is ill-defined when the argument of the square root become negative for $h<-\ttau$.
The above result is suggesting somehow that although $a_{scal}(\ttau)$ admits an asymptotic expansion it is actually an ill-defined object that actually does not exist. On the other hand we have already discussed why a well-defined $a_{scal}(\ttau)$ would be a problem given that we expect the system to be ergodic because of its finite size.
One should be aware that the mapping between the corrections $a_i$'s and the expansion of $[\sqrt{\ttau+h}]$ is a remarkable and {\it rare} exact result because typically in such computations one is limited to a few of the first coefficients. It is really a special feature of the class of SG considered, while for instance the standard (so-called {\it continuous}) SG do not have this property.
Thus it is disappointing that this remarkable mapping is only giving us a negative and somewhat involved result: {\it the quantity $a_{scal}(\ttau)$ that we expect to be ill-defined has the same well-defined expansion of the ill-defined object $[\sqrt{\ttau+h}]$!}

In order to make progress we should work with an object that we expect to be well-defined also in a finite-system size. This can be done by going back to our original dynamical problem \cite{Rizzo14}.
Indeed we expect that if we consider the dynamical expansion (\ref{Cexp}) over an appropriate {\it large but not infinite} time-scale it makes sense to assume that even in a finite-size system the correlator will stay close to the plateau value.  We have to be close to the critical temperature $|\tau|\ll 1$ and consider a large but not infinite time scale $\tau_\beta \propto 1/|\tau|^{1/(2a)}$ in order to see that $C_0(t)$ remains close to the plateau both {\it above} and {\it below} $T_c$.
The technical details of this analysis \cite{Rizzo14} are largely unpublished but the result boils down to the following scaling forms for the leading order term: 

In this regime it is well-known that the leading order term can be written in terms of Goetze's scaling function $g_0^{\pm}(t)$
\beq
C_0(t) = C_p + \sqrt{\tau} g_0^{\pm}(t/\tau_\beta)+o(\sqrt{\tau})\, \ \ t =O(\tau_\beta), \tau \ll 1  
\eeq
Similarly one can consider the finite-size corrections in (\ref{Cexp}). The technical details of this analysis \cite{Rizzo14} are largely unpublished but the result boils down to the following scaling forms  in terms of appropriate scaling functions $g_i^{\pm}(t)$
\beq
C_i(t)=\tau^{-2 i+1/2} g_i^{\pm}(t/\tau_\beta)+o(\tau^{-2 i+1/2})\, \ \ t =O(\tau_\beta), \tau \ll 1  
\eeq
The $\pm$ index stems from the fact that the scaling functions are different above ($-$) and below ($+$) the transition. Furthermore we have obviously  
\beq
\lim_{t \rightarrow \infty}g^+_i(t)=a_i
\eeq
while the corresponding limit is infinite in absolute value for the scaling functions $g_i^-(x)$ describing the region above the critical temperature.

The scaling functions $g_i^{\pm}(t)$ are rather complex objects. In order to discuss them
let us start with the leading order term $g_0^{\pm}(t)$.
It was discovered long ago \cite{Kirkpatrick87c,Crisanti93}, (see also \cite{calta1} and \cite{Parisi13}, for a recent discussion) that it obeys the very same equation of the so-called critical correlator of MCT \cite{Gotze85}
\beq
\pm 1=g^{\pm}_0(s)^2\left(1-\lambda\right) +\int_0^s (g^{\pm}_0(s-s')-g^{\pm}_0(s))\dot{g}^{\pm}_0(s')ds'
\label{SVDYN2}
\eeq
For small values of $s$ both the functions $g^{\pm}_0(s)$ diverges as $1/s^a$, while for large values of $s$ $g_0^+(s)$ goes to a constant while $g_0^-(s)$ diverges as $-s^b$ where the exponents $a$ and $b$ are determined by the so-called parameter exponent $\lambda$ according to:
\beq
\lambda={\Gamma^2(1-a)\over\Gamma(1-2a)}={\Gamma^2(1+b)\over\Gamma(1+2b)}
\label{lambdaMCT}
\eeq
A very important property is that the equation for $g^{\pm}_0(t)$ is scale invariant, in other words if we write $g_0^{\pm}(t/\tau_\beta)$ the time scale $\tau_\beta$ is undetermined. Note also that  $g_0^{\pm}(t)$ is singular at small time where it diverges. These properties are common also to the higher order corrections functions  $g_i^{\pm}(t)$: {\it they possess the same scale invariance of the leading order term and are also singular at small times}.

Now we can proceed as before and argue that it should exist a regime in which all the corrections are of the same magnitude. In order to do this we have to approach the critical temperature on a scale that depends on $N$ as we did above:
\beq
\tau=N^{-1/2} \tilde{\tau}\ .
\eeq
Then we can rewrite:
\beqa
C(t) & = & C_p+N^{-1/4}\sqrt{|\tilde{\tau}|}\left(g_0^{\pm}(t/\tau_\beta)+{g_1^{\pm}(t/\tau_\beta) \over \tilde{\tau}^2} \right.+
\nonumber
\\
&+& \left.{g_2^{\pm}(t/\tau_\beta) \over \tilde{\tau}^4}+\dots\right)+o(N^{-1/4})
\eeqa
and this suggests the existence of a unique scaling function
\beq
C(t)=C_p+N^{-1/4} g_{scal}(\ttau,t/N^{1/(4a)})
\label{dynn}
\eeq
such that the $g_i^{\pm}$'s corresponds to its asymptotic expansion for $\ttau \rightarrow \pm \infty$
\beq
g_{scal}(\ttau,\tlet) \sim \sqrt{|\tilde{\tau}|}\left(g_0^{\pm}(\tlet)+{g_1^{\pm}(\tlet) \over \tilde{\tau}^2}+{g_2^{\pm}(\tlet) \over \tilde{\tau}^4}+\dots\right)
\eeq

Note that the form (\ref{dynn}) does not imply that the correlator will remain close to the plateau value forever, this will be only true on a large but finite time scale $\tau_\beta=O(N^{1/(4a)})$, beyond which the correlator will decay to the paramagnetic/liquid value. Therefore expression (\ref{dynn}) does not imply ergodicity breaking at variance with its static counterpart (\ref{fiq}) and it is therefore consistent with the fact that a finite size system should be ergodic.

We are now in position to state the first result of \cite{Rizzo14}, {\it i.e.} the equivalence of the  $g_i^{\pm}(t)$ to SBR , through a statement similar to the one made before for the static case.
Before doing so let me first introduce the so-called simplified SBR model by considering the following dynamical stochastic equation:
\beq
\sigma  + s =-\lambda \, g^2(t)+{d \over dt}\int_0^t g(t-s)g(s)ds
\eeq
where the field $s$ is a {\it time-independent} random fluctuation of the separation parameter, Gaussian and delta-correlated in space:  
\beq
[s]=0\, ,\ [s^2]=\Delta \sigma^2 
\eeq
The above equation must be solved with a (conventional) uniform initial condition $g(t) \propto 1/t^a$ at small times. 
The simplified SBR correlator is obtained as the average solution of the stochastic equation over the random fluctuations:
\vspace{10px}
\beq
g_{sSBR}(\sigma,t) \equiv  [g(t)]
\eeq
The simplified SBR equation is nothing but a version of the full SBR equations presented earlier with no space dependence. The model in itself possess a number of interesting properties discussed in \cite{Rizzo14} and more exstensively in \cite{Rizzo14b}.
With above the definition we can state the following \cite{Rizzo14}
\begin{itemize}
\item {\it For all i the dynamical coefficients $g_i^{\pm}(t)$ are the same of the asymptotic expansion of $g_{sSBR}(t)$:}
\beq
g_{scal}(\ttau ,t) \sim g_{sSBR}(\ttau ,t)
\eeq
\end{itemize}
Note that while for the full SBR equations we cannot exhibit an explicit solution and we must resort to numerics, for the simplified model we can write down the explicit solution in terms of the functions  $g_0^{\pm}(t)$:
\beq
G_{sSBR}(\sigma,t)= \int_{-\infty}^{+\infty} {d s \over \sqrt{2 \pi} \Delta \sigma} e^{-{(s-\sigma)^2 \over 2 \Delta \sigma^2}}|s|^{1/2}g^{\sign [s]}_0(t |s|^{1/2 a} )\ ,
\label{expsol}
\eeq  
and in order to have an idea of the complexity of the coefficients $g_i^{\pm}(t)$ of the finite-volume expansion the reader can reverse-engineer them by computing the asymptotic expansion for large absolute values of $\sigma$ (or more precisely for large absolute values of $\sigma/\Delta \sigma$)
by using the representation
\beq
 {1 \over \sqrt{2 \pi} \Delta \sigma} e^{-{(s-\sigma)^2 \over 2 \Delta \sigma^2}}=\delta(s-\sigma)+{\Delta \sigma^2 \over 2}\delta''(s-\sigma)+\dots
\eeq
Note that the expansion for large positive or negative values of $\sigma$ would be an expansion around $g_0^{+}(t)$ or $g_0^{-}(t)$ respectively.  
At this stage it should be clear that SBR is not just a nice phenomenological model. The problem of finite-size corrections in SG is an example where one can derive it from first principles and verify its validity provided the scaling quantity $N \tau^2$ is small enough.
Later we will discuss that this derivation actually depends on the fact that the finite-size corrections depends on an effective Landau theory (the GCT mentioned in the introduction) and it is this theory that it is actually equivalent perturbatively to SBR. On the other hand the relevance of GCT (and thus the mapping to SBR) goes well beyond the problem of finite-size corrections in mean-field models and allows to discuss SG in physical dimension and supercooled liquids.

We also note once again that while the random fluctuations of the temperature in the {\it static} treatment leads to the ill-defined expression $[\sqrt{\ttau+h}]$ the random fluctuations in the dynamic treatment leads to expression (\ref{expsol}) which is well defined as a function of time and $\sigma$.

\subsection{Finite-Dimensional Spin-Glass Beyond Mean-Field}

In the previous subsection we have explained how some physical quantities (the expansion coefficients $g_i^{\pm}(t)$) that we can compute explicitly in some Spin-Glass models can be also computed from the asymptotic expansion of SBR.  Furthermore we are not really interested in the expansion coefficients themselves but rather in their resummation and therefore we can forget completely the expansion and take SBR as the scaling theory valid near the critical temperature.  

In the following we want to explain that this mapping to SBR is not limited to the class of fully-connected Spin-Glass models but can be extended under some assuptions first to other mean-field Spin-Glass models, second to SG in finite dimension and most importantly to super-cooled liquids.
In order to present the arguments behind the above statements we need to explain how the physical quantities $g_i^{\pm}(t)$ can be computed in the first place.

In fully connected Spin-Glass models one can show explicitly that the correlator can be computed as an integral over an appropriate action:
\beq
C=\int dQ Q \exp[-N H [Q]]
\label{genfor}
\eeq
The above expression is formal, the order parameter $Q$ is not just a  number but a complicated object that depends on two indexes (For details see \cite{Parisi13,Rizzo14}). 
The key point is the presence of a factor $N$ in front of $H[Q]$ that implies that we can take as the leading order the value of $Q^*$ that extremizes $H[Q]$, $dH[Q^*]/dQ=0$ and build a loop expansion around this value in order to obtain the $1/N$ corrections.
When performing the computation one immediately realizes that {\it i) the corrections are singular near the critical point ,ii)  the leading divergence depends  only on the first terms beyond the quadratic ones in the expansion of $H[Q]$ near $Q^*$ (in our case the cubic terms) iii) the fact that these terms have the symmetry of the original action $H[Q]$}.
The above statements amount to say that  all critical behavior is determined by the fact that independently of higher order corrections $H[Q]$ can be replaced by the following effective theory (referred as glassy critical theory (GCT) in the following ) that we identify the Landau theory of the Glass transition \cite{Parisi13,Rizzo14}:

\beqa
{\mathcal L} & = & {1\over 2}\left(- \tau \int d1 d2 \,Q(1,2)+\right.
\nonumber
\\
& + & \left. m_2\int d1 d2 d3
\,Q(1,2)Q(1,3)+m_3\int d1 d2 d3 d4 \,Q(1,2)Q(3,4) \right) 
\nonumber
\\
& - &  {1 \over
  6}w_1 \int d1 d2 d3\, Q(1,2)Q(2,3)Q(3,1)-{1 \over 6}w_2
\int d1 d2\, Q(1,2)^3
\eeqa

Now for a generic mean-field model we can argue that the correlator is still described by a form like 
(\ref{genfor}) but we may not know quantitatively $H[Q]$ and thus be unable to determine $Q^*$. Still, if we can prove (or assume) that the symmetry of $H[Q]$ is the same of our original problem it follows that they will have the same finite-size corrections (except for a rescaling of the coupling constants) because they will be controlled by the same GCT defined above.  
The above arguments implies that the mapping of finite-size corrections to SBR is relevant for all mean-field SG models that are expected to have the same type of dynamical transition of the fully connected ones. These includes {\it e.g.} Potts and Ising Spins, models defined on random lattices either of fixed or random connectivity and many others.

In the case of systems in finite dimension we have to consider an order parameter $Q(x)$ characterized by an Hamiltonian $H[Q]$. In this case there is no large factor $N$ in front of the Hamiltonian and the loop expansion is generated by computing corrections around the mean-field value that extremizes $H[Q]$. These corrections are not small because there is no large $N$ factor and once more we face the problem of resumming them.
However once again near a critical point the full Hamiltonian $H[Q]$ can be replaced by a Landau theory that determines its critical behavior.
For any system by applying some coarse graining we can always approximate the Hamiltonian as a Gaussian term plus corrections. This object is then studied by means of the loop expansion and the loop expansion may then require an appropriate renormalization treatment to be studied near the critical temperature. 
In the case of SG one can define models that leads naturally to the  GCT defined above.
In this case one can again prove the equivalence between the loop expansion of GCT and that of SBR. Therefore it is clear that whatever the procedure we have to apply to the loop expansion ({\it i.e.} renormalization) it is equivalent to study directly SBR.

\subsection{Supercooled Liquids: The Time-Scale invariance Principle}

In the case of supercooled liquids we can proceed in a similar way.
We {\it assume} that there is a critical point at some temperature and then we face the problem of selecting the most appropriate effective theory to describe it.
If GCT is the appropriate Landau theory one for a given super-cooled liquid then the mapping to SBR is granted.
Let us first note that one should not be surprised at all that two systems that are so different at the microscopic level could be described by the same Landau theory and have therefore the same critical behavior. An important example is the liquid-vapour transition that is described by the same effective theory (the so called $\phi^4$ theory) of the Ising ferromagnet and as a consequence has the same critical exponents.

The possible connection between SG and supercooled liquids was put forward almost thirty years ago based on the fact that mean-field SG's display an entropy crisis. Later it was discovered \cite{Kirkpatrick87c} that this systems exhibit a critical point similar to that of MCT, and in the following I will discuss under which assumptions it is possible to link MCT to GCT. MCT tells us that there is a transition and now we face the problem of finding the correct effective theory associated to this transition.
We know that at the leading order the function $g_0^{\pm}(t)$ are the same in MCT and in GCT and also that one can formulate a replicated version of MCT that is close to the static treatment of SG \cite{Szamel10,Rizzo13}. This is important but clearly it is not enough: we should be able to prove that MCT has corrections given by the $g_i^{\pm}(t)$'s. Unfortunately it is not clear how to improve systematically on MCT in order to compute corrections to it and thus this program is at present unfeasible.
The critical equation of MCT could be used to guess the structure and symmetry of the Hamiltonian of the associated theory but again this is not completely safe in this case. Note that MCT is a sort of black box from which certain quantities can be computed but not others, for instance at present there is no scheme to compute the four-point functions that should represent the bare propagators of the theory.

In spite of these difficulties an argument connecting MCT and GCT and thus SBR comes from a simple physical principle.
An important property of the MCT critical equation is the fact that it is invariant under a dilatation of time. Technically this manifest itself in the fact that time derivatives of the order parameter appears in appropriate combinations that are time-scale invariant.
Now if we require that the Landau theory associated to MCT is also time-scale invariant we are led naturally to the GCT.  Indeed we are looking for a theory of an order parameter $Q(1,2)$ that depends on two time indexes and time-scale invariance requires then such a theory should not depend on time derivatives of the order parameter.
On the other hand it is the presence of these time derivatives that connects different instants in time and thus guarantees causality and a unique time ordering.
As a consequence if we remove them  we are left with a theory where formally all instants in time are interchangeable, in the sense that {\it the theory is invariant under a permutation of the time indexes}. 
 Then we understand why such a theory is formally identical to a replica-symmetric theory as discussed in \cite{Parisi13}. In spite of the fact that the equations are the same the key difference is that in the dynamical theory we can look for time-dependent casual solutions thus restoring {\it a posteriori} a time-ordering that is totally absent in the replica approach. 
Summarizing we know that MCT at leading order is time-scale invariant, if we require that the Landau theory associate to it also time-scale invariant we are led naturally to GCT which is the most general theory satisfying this property (in the sense that there are no additional symmetries).

\section{Conclusions}

We started our discussion by considering finite-size corrections in a certain class of mean-field Spin-Glass models.
The computation of these corrections $g_i^{\pm}(t)$ requires a loop expansion of a dynamical field theory, the so-called glassy critical theory (GCT) that is a different object from SBR. The rigorous result announced in \cite{Rizzo14} is that GCT and SBR are equivalent perturbatively in the sense  that their loop expansions are the same at all orders.
The importance of this result is that  while GCT can only be studied perturbatively, SBR can also be studied non-perturbatively ({\it i.e.} by solving explicitly the stochastic equations) and one can demonstrate the avoided nature of the singularity and  many other non-perturbative effects, obtaining a rich characterization of the glass crossover \cite{Rizzo14,Rizzo14b,Rizzo15}.
Later we have argued that the relevance of this result goes far beyond finite-size systems, indeed the very same dynamical field theory controls also the behavior of finite-dimensional systems and the mapping to SBR remains valid in two and three dimensions.
In the case of SG, both mean-field and finite-dimensional, we can derive, at least in some important cases, an explicit direct connection between the microscopic Hamiltonian and  GCT but in the case of super-cooled liquids this is not possible, at least if we work with MCT. The problem is to identify the Landau theory associated to MCT and it is unsolvable at present  because there is no well-established scheme to compute systematically  corrections to MCT. 
Finally we have discussed how this problem can be answered in a different and more satisfactory way by 
 invoking a deeper physical principle.
Indeed MCT is characterized at criticality by time-scale invariance and if we require that the associated Landau theory should preserve this property then we are naturally led to GCT and thus to SBR.

\end{document}